\documentclass[journal=nalefd,manuscript=article]{achemso}
\usepackage{natbib}
\usepackage{graphicx}
\usepackage{amsmath}
\usepackage{color}
\usepackage{soul}
\setkeys{acs}{maxauthors=10, etalmode=truncate}
\setkeys{acs}{articletitle = false}

\newcommand*{\citen}[1]{%
  \begingroup
    \romannumeral-`\x 
    \setcitestyle{numbers}%
    \cite{#1}%
  \endgroup   
}

\title{Ferromagnetic and antiferromagnetic coupling of spin molecular interfaces with high thermal stability}

\author{Giulia Avvisati}
\affiliation{Dipartimento di Fisica, Universit$\grave{a}$ di Roma ``La Sapienza'', I-00185 Roma, Italy}
\author{Claudia Cardoso}
\affiliation{Centro S3, CNR-Istituto Nanoscienze, I-41125 Modena, Italy}
\author{Daniele Varsano}
\affiliation{Centro S3, CNR-Istituto Nanoscienze, I-41125 Modena, Italy}
\author{Andrea Ferretti}
\affiliation{Centro S3, CNR-Istituto Nanoscienze, I-41125 Modena, Italy}
\author{Pierluigi Gargiani}
\affiliation{ALBA Synchrotron Light Source, E-08290 Cerdanyola del Valles, Barcelona, Spain}
\author{Maria Grazia Betti}
\affiliation{Dipartimento di Fisica, Universit$\grave{a}$ di Roma ``La Sapienza'', I-00185 Roma, Italy}
\email{maria.grazia.betti@roma1.infn.it}

\date{\today}


\begin{document}

\maketitle

\begin{abstract}

We report an advanced organic spin-interface architecture with magnetic remanence at room temperature, constituted by metal phthalocyanine molecules magnetically coupled with Co layer(s), mediated by graphene. Fe- and Cu-phthalocyanines assembled on graphene/Co have identical structural configurations, but FePc couples antiferromagnetically with Co, up to room temperature, while CuPc couples ferromagnetically, with weaker coupling and thermal stability, as deduced by element-selective X-ray magnetic circular dichroic signals. The robust antiferromagnetic coupling is stabilized by a super-exchange interaction, driven by the out-of-plane molecular orbitals responsible of the magnetic ground state and electronically decoupled from the underlying metal via the graphene layer, as confirmed by ab initio theoretical predictions. These archetypal spin interfaces can be prototypes to demonstrate how antiferromagnetic and/or ferromagnetic coupling can be optimized by selecting the molecular orbital symmetry.

\end{abstract}

\maketitle

Paramagnetic molecules become potential building blocks in spintronics when their magnetic moments are stabilized against thermal fluctuations, e.g. by a controlled interaction with a magnetic substrate. Spin molecular interfaces, with preserved magnetic activity and exhibiting magnetic remanence at room temperature (RT) can open the route to engineer highly spin-polarized, nanoscale current sources. The need to fully control the organic spin interface and the tuning of ferromagnetic (FM) or antiferromagnetic (AFM) coupling to achieve a stable conductance has motivated a vast experimental interest~\cite{Christou_MRS_2000,Wende_NatMat_2007, Annese_PRB_2011, Annese_PRB_2013, Brede_PRL_2010}.  

In this work, we propose to optimize the thermal stability and the magnetic coupling of molecular systems, while preserving their electronic properties~\cite{Scardamaglia_JPCC_2013,Lisi_JPCL_2015}, by exploiting interlayer exchange coupling within an advanced organic spin-interface architecture: arrays of metal phthalocyanine (MPc, M=Fe,Cu) arranged on Co layer(s) intercalated below graphene~\cite{Bazarnik_ACSnano_2013,Avvisati_JPCC_2017,Avvisati_ASS_2017}.  

Herewith we demonstrate how the super-exchange interaction can be mediated by the organic ligands and the graphene layer, preserving the magnetic state of the molecule and favoring a tunable FM or AFM coupling with Co layer(s), as deduced by X-ray magnetic circular dichroism (XMCD) measurements and confirmed by state of the art theoretical predictions, unveiling the extreme sensitivity of the super-exchange interaction to the symmetry of the orbitals responsible for the magnetic state. 

Magnetic properties of molecular systems on metal surfaces can be readily modified by the molecular packing or by the orbital intermixing with the metallic states~\cite{Klar_PRB_2013}, which lead to the suppression of their local magnetic moment~\cite{Gargiani_PRB_2013, Annese_PRB_2013}. Thanks to the electronic decoupling of the graphene layer, the symmetry of the molecular orbitals carrying the magnetic moments is preserved and can be selected to optimize and stabilize the magnetic configuration of the MPcs. A major challenge is to stabilize these spin interface prototypes against thermal fluctuations, in order to achieve RT remanence. We prove that the super-exchange mechanism can induce a sizable magnetization of FePc coupled with Co layers, mediated by graphene, with a residual coupling even at RT. On the other side, the hindered superexchange path, as found for CuPc coupled with Co, results in a ferromagnetic order with reduced thermal stability. Furthermore, the easy magnetization axis of a single layer of Co intercalated under graphene is out-of-plane and switches in-plane when more than 4-5 layers are intercalated~\cite{Yang_NanoLetters_2016}. Switching the Co magnetic moment by increasing the Co thickness amplifies the magnetic response of the FePc with aligned magnetic axis, unveiling high thermal stability and remanence at RT, while the coupling with magnetically unaligned CuPc molecules is frustrated~\cite{LodiRizzini_SurfSci_2014}. 

\begin{figure*}
\centering
\includegraphics[width=1\textwidth]{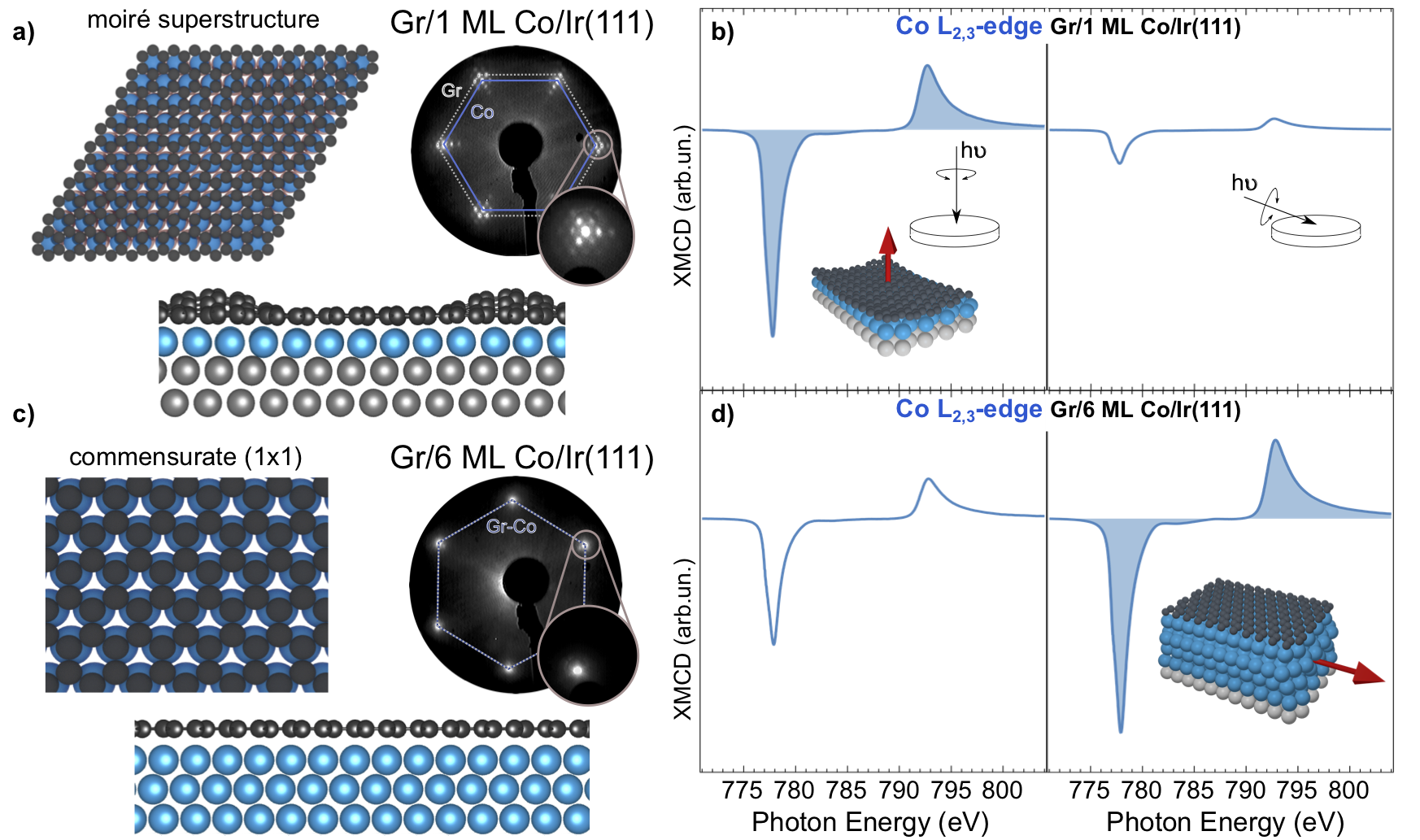}
\caption{Moir\'e superstructure of Gr on Ir(111) upon intercalation of 1 ML Co (a) and flat and commensurate Gr/Ir upon 6 ML Co intercalation (c), as deduced by ab initio DFT calculation at the LDA level and confirmed by the LEED patterns. XMCD spectra of Co L$_{2,3}$ absorption edges for Gr/1 ML Co and Gr/6 ML Co acquired in remanance at RT in normal and grazing incidence geometries. Magnetic anisotropy switches from perpendicular (Gr/1 ML Co, b) to parallel (Gr/6 ML Co, d) to the surface plane.
\label{Cobalt}}
\end{figure*}

The  magnetic (and structural) configuration of Co layer(s) intercalated under graphene can be tuned as a function of Co thickness~\cite{Yang_NanoLetters_2016}. The structural evolution of the Gr sheet upon intercalation from 1 to 6 ML of Co is consistent with what reported in Refs.~\citen{Decker_PRB_2013,Pacile_PRB_2014}, giving a picture confirmed by the theoretical predictions taking into account the complete moir\'e unit cell. When a single layer is intercalated, the Co atoms arrange pseudo-morphically to the Ir(111) surface, without altering the periodicity and symmetry of the Gr moir\'e superstructure (Fig. 1a) while its corrugation is enhanced. Further Co intercalation induces the relaxation of the lattice mismatch and the Co film recovers the bulk Co(0001) arrangement, almost commensurate with the Gr lattice, as sketched in Fig. 1c. The $1\times 1$ hexagonal pattern confirms that the Gr is commensurate and flat on the Co film. Density functional theory (DFT) simulations, taking into account the full moir\'e unit cell, fully reproduce these structural configurations. The geometry obtained for the IrCoGr 9$\times$9/10$\times$10 system shows, at the LDA level, a 1.4 \AA~corrugation for the graphene moir\'e superstructure with a minimum graphene-Co distance of 1.90 \AA \cite{PBE_D2}, Fig.~\ref{Cobalt}a. Similar findings have also been reported in Ref.~\citen{Decker_PRB_2013}. At variance, the graphene-Co distance computed for the commensurate 1$\times$1 interface is found to be 2.05~\AA{} at the LDA level~\cite{PBE_D2}, Fig.~\ref{Cobalt}c.

Turning to the magnetic properties, the XMCD spectra at the Co L$_{2,3}$ edge for a single Co layer and a Co film intercalated under the Gr sheet, obtained as the difference between the absorption edges acquired with left- and right-circularly polarized radiation, are reported in remanence condition, i.e. with no applied external field, at room temperature in the right panels of Fig.~\ref{Cobalt}. The higher dichroic response with photon impinging at normal incidence (NI) unravels the out-of-plane magnetic anisotropy of the Co layer in the incommensurate configuration with a stretched Co-Co distance, Fig.~\ref{Cobalt}b, in agreement with Refs.\citen{Decker_PRB_2013, Vita_PRB_2014}. Conversely, the higher dichroic response for Co L$_{2,3}$ XMCD at grazing incidence (GI) for the thicker intercalated Co film, reveals a magnetic state with the Co bulk in-plane easy magnetization axis. The Gr/Co heterostructures, exhibiting a tunable easy magnetization axis direction, are ideal templates to test the magnetic coupling of paramagnetic flat-lying FePc and CuPc molecules.˜\cite{Avvisati_ASS_2017}

At the early deposition stages, MPcs adsorption on graphene can be driven by the corrugated moir\'e superstructure~\cite{Mao_JACS_2009, Bazarnik_ACSnano_2013, Avvisati_JPCC_2017}. MPc molecules deposited on Gr/1ML Co/Ir are trapped in the valley regions~\cite{Avvisati_JPCC_2017} driven by the lateral electric dipole, caused by the local contraction and expansion of the Gr lattice~\cite{Zhang_PRB_2011}, and hence order in a Kagome lattice~\cite{Mao_JACS_2009, Bazarnik_ACSnano_2013}. The total energy landscape, as deduced by our DFT simulations, is almost identical for FePc and CuPc, displaying similar distances with respect to the Gr layer (3.25~\AA{} from LDA, 3.10~\AA{} from PBE-D2) and similar flat configurations upon adsorption.  Since the  adsorption geometries for  FePc and CuPc on Gr/Co/Ir are equivalent, the magnetic state can only be determined by the symmetry of the molecular orbitals involved in the magnetic coupling with the extended states of Co-intercalated graphene~\cite{Avvisati_JPCC_2017, Avvisati_ASS_2017}. 

In this respect, FePc and CuPc are paradigmatic to unveil the role of the symmetry of the active molecular orbitals in the magnetic coupling: FePc has a mostly in-plane (IP) intrinsic easy magnetization axis, while CuPc shows an out-of-plane (OOP) magnetic anisotropy~\cite{Bartolome_MPc, Stepanow_PRB_2010}.  The FePc molecule has a $d^6$ electronic ground state with S=1 spin, due to the $b_{2g}^2 e_{g}^3 a_{1g}^1 b_{1g}^0$ configuration with half-filled $e_g (d_{xz,yz})$ and $a_{1g} (d_{z^2})$ orbitals. On the other hand, CuPc molecules, with a $d^9$ ground state and S=1/2, only have a half-filled $b_{1g} (d_{x^2-y^2})$ orbital with a strongly anisotropic magnetic state perpendicular to the molecular plane~\cite{Bartolome_MPc}.   

\begin{figure*}
\centering
\includegraphics[width=1\textwidth]{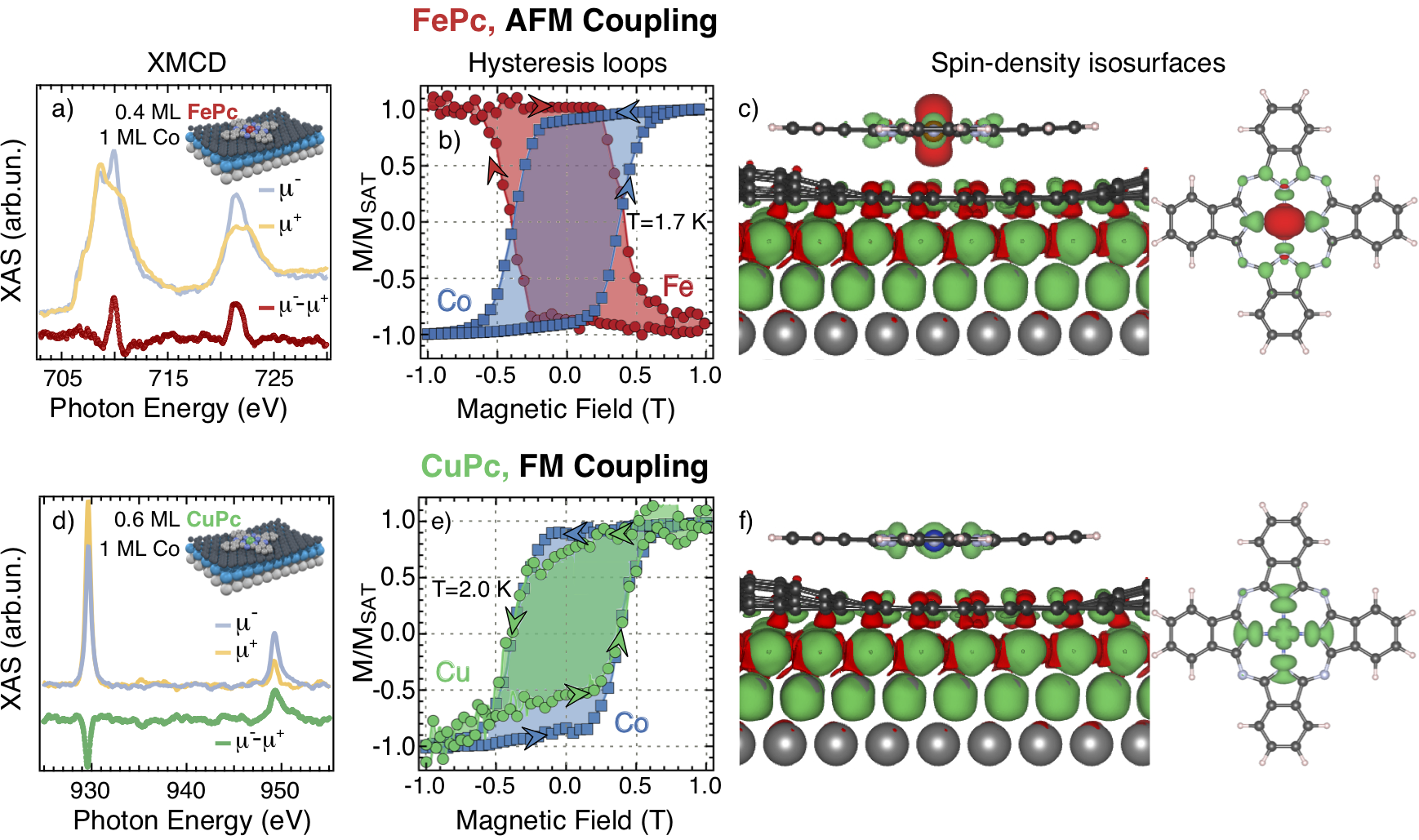}
\caption{ XMCD from Fe and Cu L$_{2,3}$ absorption edges (a,d), element-resolved hysteresis loops of FePc (b) and CuPc (e) on Gr/1ML Co. The change in the sign of the XMCD (a,d) and of the field-dependent  magnetization (b,e) indicates an AFM alignment for FePc/Gr/Co and a FM one for CuPc/Gr/Co. Spin-density isosurfaces plots for FePc and CuPc on Gr/Co/Ir (side and top view with hidden substrate), computed  at the DFT-PBE+U level, U=4 eV (c,f). The geometry optimization included PBE-D2 van der Waals corrections. Green (red) isosurfaces correspond to the up (down) spin density. 
\label{XMCD_MPc}}
\end{figure*}

In Fig.~\ref{XMCD_MPc} we report the XAS and XMCD spectra at the L$_{2,3}$ absorption edges of Fe and Cu, measured respectively at 1.7 and 2.0 K, for FePc and CuPc on Gr/1ML Co (Fig.~\ref{XMCD_MPc}a,d), having an OOP easy magnetization axis. FePc molecules couple antiferromagnetically with the intercalated Co layer, as revealed by the sign switching in the element-sensitive XMCD signal (Fig.~\ref{XMCD_MPc}a,b) with respect to the bare Gr/OOP Co (Fig.~\ref{Cobalt}b). On the other hand, CuPc molecules couple ferromagnetically when deposited on Gr/1ML Co (Fig.~\ref{XMCD_MPc}d,e). All the XAS and XMCD spectra in presence (absence) of magnetic field are reported in the Supporting Information to clearly describe the role of FePc and CuPc molecular orbitals in the magnetic state. It is worth noting that the XAS and XMCD lineshapes are preserved, except for a slight change in the relative intensities of the spectral features, for the FePc and CuPc molecules upon adsorption on Gr/Co. Thus, the spin and orbital contribution to the magnetic moments, as deduced applying the sum rules to the XMCD signals˜\cite{Thole_PRL_1992,Carra_PRL_1993} and confirmed by theoretical predictions, are comparable with what reported for the free-standing molecule and/or molecular films˜\cite{Bartolome_PRB_2010,Lisi_JPCL_2015,Gargiani_PRB_2013}, confirming the decoupling role of the graphene layer.

Magnetism generally emerges from short-ranged interactions between molecular units due to the localization of the atomic wave functions. To achieve a magnetic interaction over a large distance, a super-exchange mechanism can be invoked, where the bridging over non-magnetic organic ligands mediates higher-order virtual hopping processes. In metal phthalocyanines, the organic ligands can mediate the coupling between the central transition metal ions and a magnetic layer~\cite{LodiRizzini_SurfSci_2014, Bernien_PRL_2009, Klar_PRB_2013, Hermanns_AdvMat_2013}. In our spin interface architecture a Fe-N-Gr-Co super-exchange path can be driven by a weak hybridization between the $\pi$ orbitals of the pyrrolic N atoms and the $e_g (d_{xz,yz})$ and $a_{1g} (d_{z^2})$ orbitals of the central Fe ion. In the CuPc molecule the orbitals protruding out of the molecular plane are completely filled and the magnetic moment is mostly carried by a single planar orbital, therefore the proposed superexchange interaction path is hindered and the interaction switches to FM.

The confirmation of the AFM and FM coupling found experimentally and the role of the molecular orbitals involved in the magnetic response are clarified by DFT calculations. In Fig.~\ref{XMCD_MPc}(c,f) we report the computed spin densities for FePc and CuPc on a Gr/1ML Co/Ir at PBE+U level, with the spin up (down) iso-surfaces represented in green (red). The Fe-N-Gr-Co super-exchange path is determined by a spin imbalance located at the central ion and at the surrounding N and C atoms. The spin of the central ion is oriented opposite to the one of N and C, and anti-ferromagnetically coupled with the underlying Co spin moment. In the case of CuPc, the spin imbalance is located at the central ion and on the surrounding N atoms, and coupled ferromagnetically with the Co spin moment. Top views of the spin densities (panels c,f) better display the shape of the orbitals involved and the flips of the local magnetic moments. Quantitative data (atom-resolved magnetic moments, as given by L\"owdin charge analysis) are provided in the Supporting Information.

A detailed analysis of L\"owdin charges can also be used to further discriminate the symmetry of the molecular orbitals involved in the magnetic coupling. In the case of CuPc, upon adsorption the magnetic moment of the molecule is mostly carried by the half-filled $d_{x^2-y^2}$ orbital of the Cu ion, corresponding to the $b_{1g}$ state of CuPc free molecule, highly hybridized with the pyrrolic N atoms of the macrocycle~\cite{Rosa_InorgChem_1994}, as evident in Fig.~\ref{XMCD_MPc}(f).
The situation is slightly more complicated for FePc, where at least two or three Fe-$d$ orbitals are involved. In particular, as shown in Tab. S3 in Supporting Information, the magnetic moment is carried by $d_{z^2}$ and either one or both $d_{xz}$ and $d_{yz}$ (depending on the functional, see Supporting Information for the detailed discussion), corresponding to the $a_{1g}$ and $e_g$ orbitals of free standing FePc.
The above picture is very robust against the use of different exchange and correlation functionals (LDA, PBE, PBE+U, discussed in the Supporting Information), and is in excellent agreement with the experimental data, fully supporting the picture where the magnetic coupling of MPc with Gr/Co/Ir is mainly driven by a super-exchange channel, selected by the symmetry of the involved molecular orbitals. Indeed, we herewith report on two super-exchange paths, both actively mediated by the Gr sheet and the molecule organic backbone, inducing either an AFM (180$^{\circ}$ super-exchange, FePc) or a FM (90$^{\circ}$ super-exchange, CuPc) coupling at the spin interface. These mechanisms dominate the magnetic interaction, as deduced by the increase of the magnetic moment projected on the C and N atoms of the molecule macrocycle upon adsorption (see Tab. S2 in the Supporting Information file). Nevertheless, a direct interaction between the Fe $d_{z^2}$ state and the Gr $\pi$ orbitals, AFM coupled with the intercalated Co layer~\cite{Decker_PRB_2013,Vita_PRB_2014} and hence reinforcing the antiparallel spin alignment, cannot be excluded.

\begin{figure}
\centering
\includegraphics[width=0.5\textwidth]{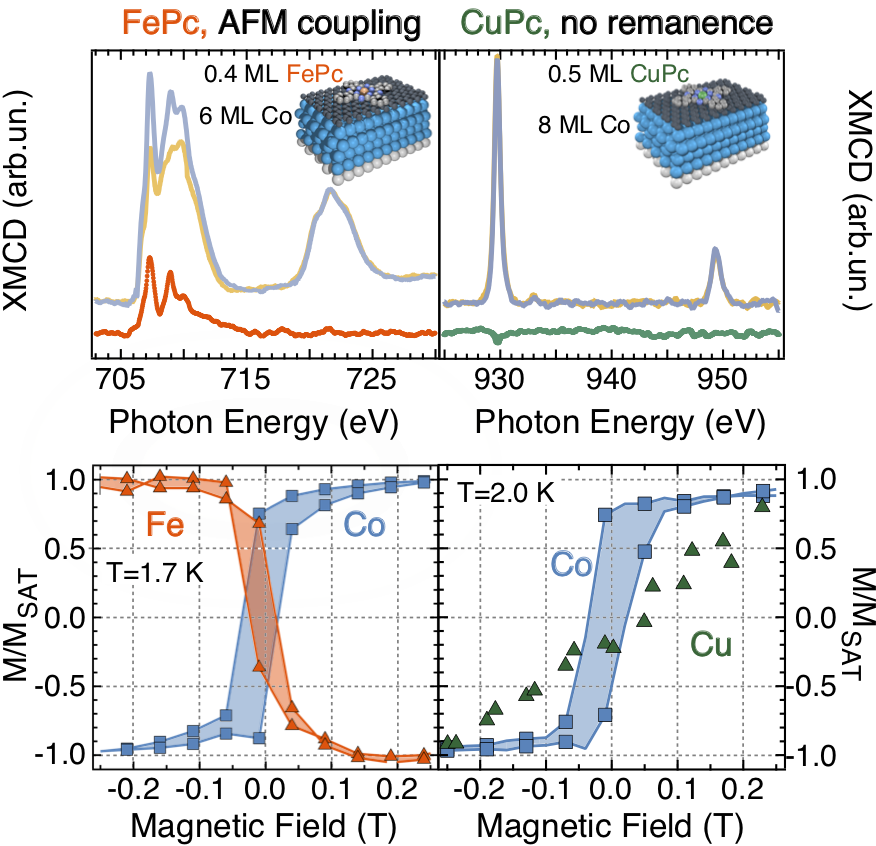}
\caption{XMCD from Fe and Cu  L$_{2,3}$ absorption edges for FePc (left) and CuPc (right) molecules deposited on Gr/IP Co in the upper panels. Hysteresis loops are presented in the lower panels, confirming that the coupling is AFM for FePc/Gr/6 ML Co, while the CuPc has no remanence.
\label{MPc_GrIPCo}}
\end{figure}

A further step forward in the optimization of this spin-interface is to enlighten the role of the relative orientation of MPc/Co easy magnetization axes on the stability of the magnetic coupling. FePc and CuPc molecules adsorb flat-lying on the commensurate Gr on the Co film with in-plane magnetization. XMCD spectra for FePc  and CuPc, respectively on Gr/6 ML Co and Gr/8 ML Co, are reported together with the field-dependence of the normalized magnetization in Fig.~\ref{MPc_GrIPCo}. While an AFM coupling is confirmed for FePc/Gr/Co when the easy magnetization axes are aligned (left panel), the CuPc molecules adsorbed on the flat graphene layer on Co has a negligilble XMCD signal at zero magnetic field (right panel). The magnetic coupling is more effective when the magnetization easy axis of the Gr-covered Co layer(s) and the one of the molecules are aligned, i.e. when the FePc molecules are adsorbed on the thick Gr/Co film and the CuPc on the single Gr/Co layer, regardless on the origin of the magnetic interaction. More details on the field- and direction-dependence of the CuPc and FePc magnetic state are reported in the Supporting Information.

\begin{figure}
\centering
\includegraphics[width=0.5\textwidth]{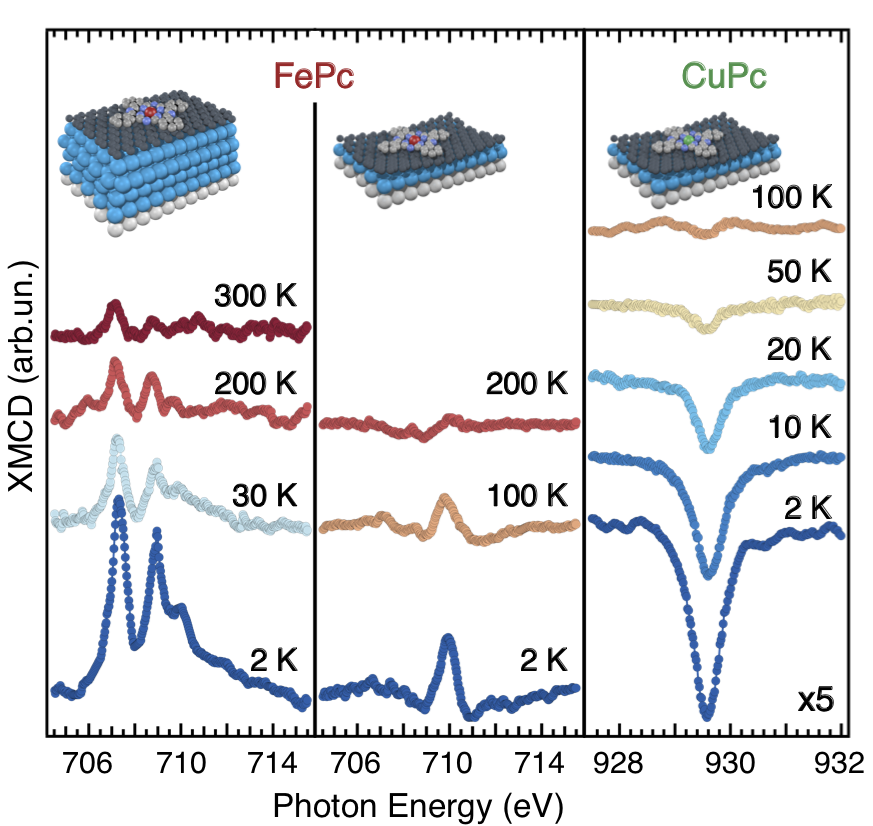}
\caption{Thermal evolution of the XMCD signal for FePc/Gr on IP (left) and OOP (middle) Co and CuPc/Gr on OOP Co (right).
\label{thermal_MPc}}
\end{figure}

A fine tuning of these spin interface architectures requires to evaluate the robustness of the magnetic response against thermal fluctuations, thanks to temperature-dependent XMCD measurements (Fig.~\ref{thermal_MPc}). It is worth to point out that all measurements were acquired in absence of a magnetic field and over several hours, indicating a very long relaxation time and a stable magnetic configuration of the coupled MPc-Gr-Co systems. 

Indeed, the super-exchange path stabilizes the AFM coupling up to 200 K for FePc/Gr/OOP Co, while for FePc/Gr/IP Co a residual XMCD signal is  detectable even at RT. The FePc magnetic state coupled to Gr/OOP Co and Gr/IP Co can be understood in terms of the relative orientation between the molecule and substrate easy magnetization axes, since an extra amount of energy is needed to turn the FePc magnetization, from its intrinsic IP direction, to adapt to the OOP Co easy magnetization axis. The magnetic-coupling for CuPc/Gr/OOP Co is much weaker, being not driven by a 180$^{\circ}$ super-exchange mechanism, as reflected by the disappearance of any XMCD remanent signal above 50 K. These differences can be quantitatively evaluated by performing a Brillouin fit over the thermal evolution of the XMCD intensity, leading to a coupling energy of 2.8 $\pm$ 0.6 meV and 2.1 $\pm$ 0.5 meV for FePc ong Gr/1 and 6 ML Co, respectively, while the less stable CuPc/Gr/1 ML Co system exhibits a much lower coupling energy of 0.6 $\pm$ 0.2 meV (for detailed discussion see the Supporting Information file).

A direct magnetic coupling has been also detected for FePc molecules adsorbed on Gr grown on Ni(111), leading to a ferromagnetic alignment˜\cite{Candini_JPCC_2014}, while in-plane AFM coupling mediated by graphene was recently observed between a Ni thin-film and Co-porphyrin molecules up to 200 K~\cite{Hermanns_AdvMat_2013}.  Despite a comparable exchange energy, the magnetic stability against thermal fluctuations up to room temperature is here observed for the first time. In these peculiar spin interfaces the Gr layer plays a dual role, on one hand it acts as a buffer layer, inhibiting the MPc-Co electronic interaction and preserving the magnetic properties of the adsorbed molecules, while, on the other hand, it actively participate to the magnetic coupling, either by a direct coupling˜\cite{Candini_JPCC_2014} or by opening an effective super-exchange channel.~\cite{Hermanns_AdvMat_2013} 

In conclusion, these graphene-spaced spin molecular interfaces present either AFM or FM coupling driven by different super-exchange paths, as supported by the the ab-initio theoretical calculations and the experimental XMCD results. Despite the almost identical structural configurations and a large distance between the Fe(Cu) metal centers and the magnetic Co substrate, the magnetic coupling is very robust against thermal fluctuations. In particular, the Fe-N-Gr-Co super-exchange channel drives an AFM coupling, favored by the presence of out-plane molecular orbitals, mediated by the organic ligands and graphene. On the other side, in the Cu-N-Gr-Co the magnetic coupling is strongly weakened by 90$^\circ$ super-exchange interaction path leading to a FM coupling of the molecule to the Co layer. This scenario, completely supported by the theoretical spin density calculated on the whole moir\'e cell, is robust against different exchange and correlation functionals. The choice of an effective super-exchange path can ensure  the stability against thermal fluctuations, even at RT, and it can be further optimized with a fine control of the relative orientation of the easy magnetization axes at the spin interface. In perspective, the magnetic remanence at RT of these archetypal spin interfaces, once paired with a tunable magnetic substrate, opens the possibility to produce future operational spintronic devices.

\section*{Experimental and computational methods}
\textit{Sample preparation:} The Ir(111) surface was prepared with several cycles of 2 keV Ar$^+$ sputtering, followed by annealing above 1300 K. The Gr sheet was grown on the clean and ordered Ir(111) surface by exposing the substrate to a partial pressure of 10$^{-6}$ mbar of ethylene (C$_2$H$_4$), and annealing the covered surface above 1500-1600 K. The completion of a single domain Gr layer was confirmed by the presence of sharp and bright spots of the moir\'e superstructure in the LEED diffraction pattern. Metallic Co was then sublimated with an e-beam evaporator and deposited on Gr/Ir(111) kept at room temperature. Finally, the Co/Gr/Ir(111) sample was annealed at 600-800 K to favour Co intercalation, following Refs.~\citen{Pacile_PRB_2014, Decker_PRB_2013, Vita_PRB_2014}. The quantity of intercalated Co was determined with a quartz crystal microbalance and double-checked with an Auger-calibrated growth compared with the XAS jump edge ratio (see supporting information of Ref.˜\citen{Gargiani_NatComm_2017}), highlighting a layer-by-layer growth. MPc powders (M=Fe,Cu) were sublimated with a home-made resistively heated quartz crucible, at a constant rate of 0.3 \AA/min, measured with a quartz crystal microbalance.

\textit{XAS and XMCD measurements:} X-ray absorption spectroscopy and magnetic circular dichroism (XAS and XMCD) measurements were performed at the BOREAS beamline of the Alba synchrotron radiation facility~\cite{BOREAS_bml} in total electron yield (TEY), by measuring the sample drain current normalized with respect to incident flux, measured as the drain current on a clean gold grid. The measurements were performed in two different experimental geometries, in order to probe the magnetic response along the easy as well as the hard magnetization axis. The in-plane magnetic state was studied by impinging the sample at GI, namely at 70$^{\circ}$ incidence angle; while the out-of-plane magnetic response is determined at NI. The hysteresis curves have been obtained by normalizing the field-dependent $L_3$ XMCD intensity at a pre-edge signal, in order to cancel-out any field-induced artifact.

\textit{Theoretical modeling:} Plane wave density functional theory calculations were performed using the Quantum ESPRESSO~\cite{Giannozzi_JPhysConMat_2009,gian+17jpcm} package at the LDA~\cite{Perdew_PRB_1981}, GGA-PBE~\cite{Perdew_JCP_1996} and PBE+U~\cite{Cococcioni_PRB_2005} level (with a value of U=4 eV for the d-orbitals of the MPc central ion).
Structural relaxations were carried out using LDA and PBE exchange-correlation potential, including van der Waals interactions within the semi-empirical method of Grimme (DFT-D2)~\cite{Grimme_JCompChem_2006}.
The IrCoGr system was simulated with a super-cell consisting of a IrCo slab (three Ir and one Co layers) with a 9$\times$9 lattice in-plane periodicity and an overlaying 10$\times$10 graphene layer. The same system was then relaxed with one FePc (CuPc) molecule adsorbed on graphene. 
Technical details of the calculation are reported in the Supporting Information.

\begin{acknowledgement}
These experiments were performed at the BOREAS beamline at ALBA Synchrotron with the collaboration of ALBA staff. The authors thank Pierluigi Mondelli for experimental assistance during the beamtime. AF, DV, CC acknowledge financial support from the EU Centre of Excellence ``MaX - Materials Design at the Exascale'' (Horizon 2020 EINFRA-5, Grant No. 676598). Computational resources were partly provided by PRACE (Grant No. Pra11\_2921) on the Marconi machine at CINECA. GA and MGB acknowledge fundings from Sapienza University of Rome. PG acknowledges funding from the Spanish MINECO (grant No. FIS2013-45469-C4-3-R).

\end{acknowledgement}

\begin{suppinfo}
Detailed material related to the XAS measurement at the Fe and Cu $L_{2,3}$ edges, with and without magnetic field and in the two experimental geometries, for the spin interface and to DFT modelling are available free of charge in the Supporting Information file.
\end{suppinfo}

\bibliographystyle{acs}
\bibliography{bibliography}

\begin{mcitethebibliography}{38}
\bibitem[PBE()]{PBE_D2}
The inclusion of van der Waals interactions at the PBE-D2 level does not change
  significantly the picture, with a graphene corrugation of 1.3~\AA~and a
  minimum graphene-Co distance of 1.84 \AA. At variance, the Gr-Co distance for
  the 1$\times$1 commensurate system is found to be 2.13~\AA{} for
  PBE-D2.\relax
\mciteBstWouldAddEndPunctfalse
\mciteSetBstMidEndSepPunct{\mcitedefaultmidpunct}
{}{\mcitedefaultseppunct}\relax
\EndOfBibitem
\end{mcitethebibliography}
\end{document}